\documentclass[lettersize,journal]{IEEEtran}
\usepackage{amsmath,amsfonts}
\usepackage{algpseudocode}
\usepackage{algorithm}
\usepackage{array}
\usepackage[caption=false,font=normalsize,labelfont=sf,textfont=sf]{subfig}
\usepackage{textcomp}
\usepackage{stfloats}
\usepackage{url}
\usepackage{verbatim}
\usepackage{graphicx}
\usepackage{cite}
\usepackage{blindtext}
\usepackage{amsmath, amsthm, amsfonts}
\usepackage[english]{babel}
\usepackage{enumerate}
\usepackage{mathtools}
\usepackage{balance}
\usepackage{array}
\usepackage{multirow}
\errorcontextlines\maxdimen

\makeatletter
    \newcommand*{\algrule}[1][\algorithmicindent]{\makebox[#1][l]{\hspace*{.5em}\thealgruleextra\vrule height \thealgruleheight depth \thealgruledepth}}%
\newcommand*{\thealgruleextra}{}
\newcommand*{\thealgruleheight}{.75\baselineskip}
\newcommand*{\thealgruledepth}{.25\baselineskip}

\newcount\ALG@printindent@tempcnta
\def\ALG@printindent{%
    \ifnum \theALG@nested>0
        \ifx\ALG@text\ALG@x@notext
        \else
            \unskip
            \addvspace{-1pt}
            \ALG@printindent@tempcnta=1
            \loop
                \algrule[\csname ALG@ind@\the\ALG@printindent@tempcnta\endcsname]%
                \advance \ALG@printindent@tempcnta 1
            \ifnum \ALG@printindent@tempcnta<\numexpr\theALG@nested+1\relax
            \repeat
        \fi
    \fi
    }%
\usepackage{etoolbox}
\patchcmd{\ALG@doentity}{\noindent\hskip\ALG@tlm}{\ALG@printindent}{}{\errmessage{failed to patch}}
\makeatother

\newbox\statebox
\newcommand{\myState}[1]{%
    \setbox\statebox=\vbox{#1}%
    \edef\thealgruleheight{\dimexpr \the\ht\statebox+1pt\relax}%
    \edef\thealgruledepth{\dimexpr \the\dp\statebox+1pt\relax}%
    \ifdim\thealgruleheight<.75\baselineskip
        \def\thealgruleheight{\dimexpr .75\baselineskip+1pt\relax}%
    \fi
    \ifdim\thealgruledepth<.25\baselineskip
        \def\thealgruledepth{\dimexpr .25\baselineskip+1pt\relax}%
    \fi
    \State #1%
    \def\thealgruleheight{\dimexpr .75\baselineskip+1pt\relax}%
    \def\thealgruledepth{\dimexpr .25\baselineskip+1pt\relax}%
}

\begin{document}
	
\title{AI-assisted Improved Service Provisioning for Low-latency XR over 5G NR}


\author{Moyukh~Laha,~\IEEEmembership{}
    	Dibbendu~Roy,~\IEEEmembership{}
     Sourav~Dutta,~\IEEEmembership{}
     Goutam Das~\IEEEmembership{}
    	
\IEEEcompsocitemizethanks{\IEEEcompsocthanksitem M. Laha, S. Dutta, and G. Das are with the G. S. Sanyal School of Telecommunications, Indian Institute of Technology Kharagpur, India. \\D. Roy is with School of Electrical Engineering and Computer Science, KTH Royal Institute of Technology, Sweden.\\
Corresponding author: Moyukh Laha, E-mail: laha.moyukh@ieee.org \IEEEcompsocthanksitem }
}

\maketitle

\begin{abstract}
Extended Reality (XR) is one of the most important 5G/6G media applications that will fundamentally transform human interactions. However, ensuring low latency, high data rate, and reliability to support XR services poses significant challenges. This letter presents a novel AI-assisted service provisioning scheme that leverages predicted frames for processing rather than relying solely on actual frames. This method virtually increases the network delay budget and consequently improves service provisioning, albeit at the expense of minor prediction errors. The proposed scheme is validated by extensive simulations demonstrating a multi-fold increase in supported XR users and also provides crucial network design insights. 
\end{abstract}
	
\IEEEpeerreviewmaketitle

\section{Introduction}
\label{intro}
Extended Reality (XR) encompasses immersive technologies like Virtual Reality (VR), Mixed Reality (MR), and Augmented Reality (AR), revolutionizing human interactions across various industries, e.g., entertainment, education, healthcare, etc. However, delivering XR services over 5G NR poses significant challenges in meeting strict requirements for data rate ($ \sim100Mbps$), reliability ($>=99\%$), and latency ($ \sim 2.5ms$) \cite{xr_eric}. To tackle these challenges, the 3rd Generation Partnership Project (3GPP) has proposed the \emph{split rendering architecture} for XR, where the processing is offloaded to the Edge node \cite{xr_split}. This approach involves sending data from an XR user to the Edge for processing and subsequent distribution of these to one or multiple XR users.

\par There is a paucity of research on service provisioning for low-latency XR in 5G NR. The standardization of XR services in 5G NR is being extensively investigated in release-18, also known as 5G-advanced, by the 3GPP standardization forums \cite{xr_6g}. Previous studies, such as \cite{xr_lit2}, have highlighted the shortcomings of 5G NR through system-level performance evaluations of XR applications, and in \cite{xr_10}, the authors introduced the concept of frame-level integrated transmission to enhance performance. However, none of the existing studies have thoroughly examined low-latency XR in the context of 5G NR. We now demonstrate the inadequacy of 5G NR in supporting low-latency XR, which motivates our work.

\par Table \ref{tab:perf} shows the performance of low-latency XR over 5G NR in terms of the number of satisfied XR users. A user is deemed satisfied when 99\% of its frames are delivered within the specified network delay bound \cite{xr_eric}, defined as \textit{delay reliability} of 99\%. These results were obtained through extensive simulations employing various scheduling schemes (proportionality fair (PF), deficit round robin (DRR), and maximum CQI-based scheduling (MAX-CQI) \cite{5g_scheduling}), adhering to 3GPP specifications, and utilizing Simu5G platform (see Section \ref{sec:simu} for details). The third column of the table unequivocally illustrates the inadequacy of 5G NR in meeting the XR requirements. Even with a system bandwidth of 100 MHz and a delay bound ($\tau$) of 2.5 ms—typical for cloud gaming services \cite{xr_eric}—5G NR can only support a maximum of one user per base station for the specified data and frame rates. Such a low number of \textit{simultaneous XR support} (defined as the number of satisfied XR users) is because XR applications demand not only strict low latency but also generate large XR frames periodically (in the range of hundreds of KB), resulting in data bursts that may require multiple transmission slots to clear \cite{xr_10}, which often leads to delay reliability violations. 
\begin{table}[]
\centering
\caption {No. of Delay Reliability Satisfied XR Users} \label{tab:perf}
\begin{tabular}{|c|c|cc|}
\hline
\multirow{2}{*}{\begin{tabular}[c]{@{}c@{}}Frame rate ($f_r$ in fps)\\ Data rate ($d_r$ in Mbps)\end{tabular}} & \multirow{2}{*}{\begin{tabular}[c]{@{}c@{}}Scheduling \\ Scheme\end{tabular}} & \multicolumn{2}{c|}{\begin{tabular}[c]{@{}c@{}}Delay Satisfied XR Users \\for Delay Bound $\tau$ (ms)\end{tabular}} \\ \cline{3-4} 
                                                                                                               &                                                                               & \multicolumn{1}{c|}{$\tau$ = 2.5 }                      & $\tau$ = 2.5 + 1/$f_r$                       \\ \hline
\multirow{3}{*}{60, 30}                                                                                        & PF                                                                            & \multicolumn{1}{c|}{1}                           & 21                                    \\ \cline{2-4} 
                                                                                                               & DRR                                                                           & \multicolumn{1}{c|}{1}                           & 21                                    \\ \cline{2-4} 
                                                                                                               & MAX-CQI                                                                       & \multicolumn{1}{c|}{1}                           & 21                                    \\ \hline
\multirow{3}{*}{120, 60}                                                                                       & PF                                                                            & \multicolumn{1}{c|}{1}                           & 10                                    \\ \cline{2-4} 
                                                                                                               & DRR                                                                           & \multicolumn{1}{c|}{1}                           & 11                                    \\ \cline{2-4} 
                                                                                                               & MAX-CQI                                                                       & \multicolumn{1}{c|}{1}                           & 11                                    \\ \hline
\end{tabular}
\vspace{-5mm}
\end{table}

\par This letter addresses the limitations of 5G NR in supporting low-latency XR and achieving a high number of simultaneous XR support. We introduce an AI-assisted cross-layer scheme that utilizes AI techniques to predict application layer data. This predicted data is then processed and scheduled at the MAC layer to ensure the required delay reliability. We also consider the trade-off between extending the delay budget and potential user experience degradation caused by prediction errors. Our main contributions are as follows:

\begin{itemize}
\item We propose a novel AI-assisted service provisioning scheme for low-latency XR, along with comprehensive guidelines for the necessary architectural modifications required in the edge server.
\item We provide simple yet straightforward network design guidelines for determining the prediction duration for homogeneous XR users.
\item We validate the effectiveness of our proposed scheme through extensive simulations in Simu5G, demonstrating a many-fold increase in supported XR users.
\end{itemize}

\section{AI-assisted Service Provisioning}\label{sec:main}

\subsection{The Philosophy}
\begin{figure}[]
\centering
\includegraphics[scale=0.36]{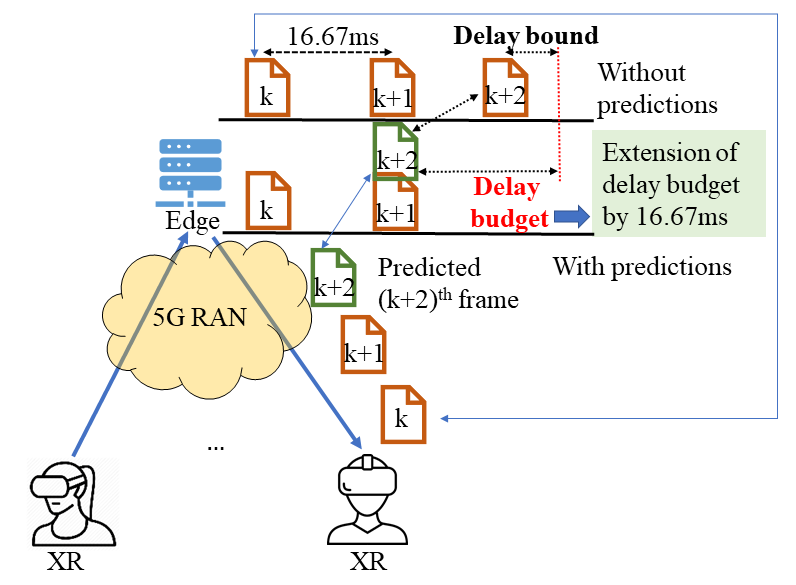}
\caption{{The figure shows by predicting a single future frame, the delay budget is increased by 16.67 ms for 60 fps XR.}}
\label{fig:intro}
\vspace{-0.5cm}
\end{figure}

The last column of Table \ref{tab:perf} reveals that relaxing the delay bound by one frame duration (e.g., 16.67 ms for a 60 fps XR) significantly increases the number of supported XR users while maintaining 99\% delay reliability. However, physically relaxing the bound is not feasible as it would lead to a poor user experience. We could, however, take a different approach. By predicting only one XR frame at the application layer, which typically contains correlated data and is, therefore, easy to predict, we can achieve a 16.67 ms increase in the network delay budget. And predicting multiple future frames with minor prediction errors is readily achievable using state-of-the-art deep learning techniques. This predicted data can then be processed and scheduled at the MAC layer within the extended delay budget to satisfy the required delay reliability at the cost of minor prediction errors. It is important to note that this approach enhances the delay budget without altering the actual delay bound, as illustrated in Fig. \ref{fig:intro}. It may further be noted that implementing this approach requires analyzing application layer data within the network, which is facilitated by the XR split rendering architecture.

\subsection{Architectural Modifications of the Edge }
\begin{figure}[]
		\centering
		\includegraphics[scale=0.33]{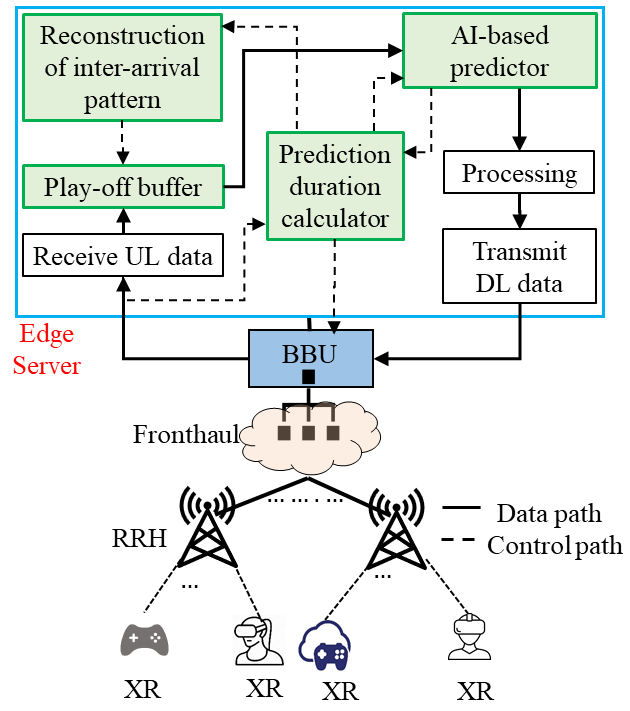}
    		\caption{{Modified architecture with system model. New additions are highlighted in green.}}
		\label{fig:system_model}
\end{figure}

Modifications to the edge node's architecture are essential to enable AI-assisted service provisioning, as depicted in Figure \ref{fig:system_model}. The newly added functional units are highlighted in green, while the existing split rendering architecture remains unchanged. It is worth noting that the frames received by the edge from XR users may experience variable delays in reaching the edge server (full details of the scheme are provided in the following subsection). Since predictions are based on these received data at the edge server, the variability in the inter-arrival pattern of these data may lead to increased prediction errors. To mitigate such errors, a play-off buffer is incorporated into the modified architecture to reconstruct the inter-arrival data pattern. The AI unit utilizes the received data to predict future frames for a specified duration (explained later) and processes them for scheduling to the XR users.

\subsection{The Algorithm with Working Principle}

The XR service involves a distinct physical process characterized as follows. XR users transmit XR frames to the Edge,  which are initially stored, and then processed at the Edge. Following processing, new frames are generated and subsequently transmitted to either the same or different XR users. To ensure the delay reliability requirements, the processed XR frames must be delivered within the specified delay bound, which is difficult to achieve, as described earlier. Our proposed scheme, outlined in Algorithm \ref{alg:app} and executed at the Edge node, introduces modifications to achieve this objective. The algorithm for user $n$ requires the following input parameters: the prediction interval denoted as $p^n_d$; the inter-frame duration $T_f^n$; and the allowed maximum delay bound $D_{UB}^n$. By processing these inputs, the algorithm generates XR frames that are designated for downlink scheduling. At the application layer of the algorithm, the edge node verifies whether the last bit of the $i^{th}$ frame $F_i^n$, denoted as $Rx_t(F^n_i)$, is received within the duration ($p^n_d T^n_f + D^n_{UB}$). If this condition holds true, the $i^{th}$ original frame $F_i^n$ is enqueued in the predictor buffer, which the predictor unit utilizes for subsequent predictions. Conversely, if $i^{th}$ frame is not received within the time interval ($p^n_d T^n_f + D^n_{UB}$), then the $i^{th}$ frame is also predicted denoted as ${F_i^n}^*$ and subsequently enqueued in the predictor buffer. Note that once the original frame is received, the predicted frame in the predictor buffer may be replaced to minimize the prediction errors. Finally, the $(i+p_d^n)^{th}$ frame is predicted based on these enqueued predictor buffer contents and forwarded to the lower layers for scheduling. At the MAC layer, as long as the MAC queue is not empty, the downlink scheduling subroutine $DLSchedule()$ is invoked. It is worth noting that our proposed scheme seamlessly integrates with any existing downlink scheduling scheme. In this work, we specifically consider three widely adopted 5G scheduling schemes: proportionality fair (PF), deficit round robin (DRR), and maximum CQI-based scheduling (MAX-CQI) \cite{5g_scheduling}.
\par
\begin{figure}[]
		\centering
		\includegraphics[scale=0.34]{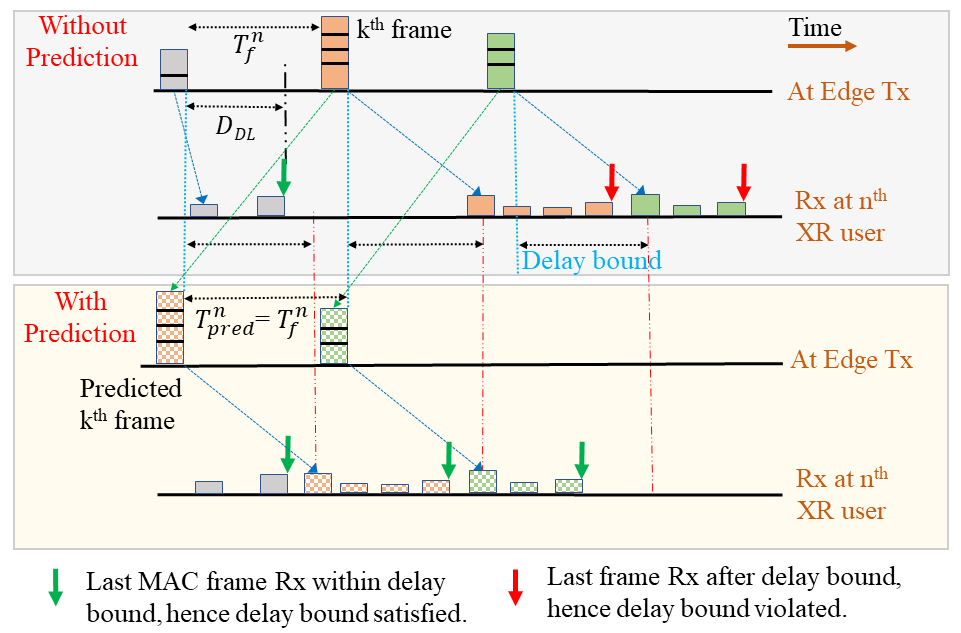}
		\caption{{The divisions of XR frames indicate its fragmentation into smaller MAC frames for transmission. Successful reception of an XR frame means its last MAC frame is received within the delay bound. Here, the delay bound of the $k^{th}$ and $(k+1)^{th}$ frames are violated if no prediction is employed. However, with the one-frame prediction (1FP, i.e., $p_d^n = 1$), the delay reliability of the frames can be met.}}
		\label{fig:motiv}
  \vspace{-0.25cm}
\end{figure}

 We now describe how the proposed scheme could better satisfy the required delay reliability. In the downlink, the frames experience a variable downlink delay, symbolized by $D_{DL}$. As described earlier, this is constrained by the allowed delay bound $D_{UB}$, typically in the millisecond range. However, achieving such low values is not easy and often gets violated, as shown in Table \ref{tab:perf}. In our proposed scheme, an AI-based predictor predicts future application-layer data (XR frames), and this predicted data is scheduled instead of the original data, as described in lines 2 to 13 of the algorithm. Since the data is predicted for prediction interval $p^n_d$, the processing and downlink scheduling could start $T^n_{pred} = p^n_d T^n_f$ duration earlier. Therefore, for the $k^{th}$ frame of the $n^{th}$ XR user, we could write the following:

\begin{align}\label{tot_delay_inequality}
	 D_{DL}^{k,n} = T_{pred}^n + D_{UB}
\end{align}
Equation \ref{tot_delay_inequality} clearly shows that introducing a predictor at the edge server to predict future data for a duration of $T_{pred}^n$ virtually increases the delay budget by the duration of $T_{pred}^n$. This is illustrated in Fig. \ref{fig:motiv}. Since the delay budget is increased, the probability of satisfying the delay reliability also increases, and consequently, a higher number of XR could be supported. Essentially this strategy is a cross-layer approach wherein application layer prediction is utilized to relax the delay budget of the MAC scheduling, which aids in meeting the overall network delay reliability at the cost of minor prediction errors.

\begin{algorithm}
\caption{Proposed Scheme: Running at Edge for user $n$}\label{alg:app}
\hspace*{\algorithmicindent} \textbf{Input:} Prediction interval $p_d^n$, Frame duration $T_f^n$, Delay bound $D^n_{UB}$.\\
\hspace*{\algorithmicindent} \textbf{Output:} XR frames for scheduling
in the downlink.
\begin{algorithmic}[1]
\State At \textbf{Application layer}:
\While{True}
\For{$i^{th}$ frame of user $n$, $F_i^n$ ({$ i \in {1,2,...}$)}}
\If{$Rx_t(F_i^n) \leq p^n_d.T^n_f + D^n_{UB}$}
\State Enqueue $F_i^n$ in predictor buffer of $n^{th}$ user
\ElsIf{$Rx_t(F_i^n) > p^n_d.T^n_f + D^n_{UB}$}
\State Predict $F_i^n$ frame as ${F_i^n}^*$
\State Enqueue ${F_i^n}^*$ in predictor buffer of $n^{th}$ user
\EndIf
\State Predict $F_{i+p_d^n}^n$ frame based on all previous enqueued frames
\EndFor
\State Send the XR frames to the MAC Queue
\EndWhile
\State At \textbf{MAC layer}:
\While{MAC Queue $!=$ NULL}

\State Call downlink scheduling subroutine $\textbf{DLSchedule}()$
\EndWhile
\end{algorithmic}
\end{algorithm}

\subsection{The AI Model}
We now present an AI-based predictor that utilizes received data at the edge node to predict future XR frames. Our approach employs a Conditionally Reversible Network for prediction, inspired by the CrevNet model \cite{crevnet}. Unlike conventional architectures that combine convolutional and recurrent layers for video frame prediction, our network utilizes reversible topologies to construct a bijective two-way autoencoder and a complementary recurrent predictor. This approach addresses the limitations of traditional models, including high memory usage and computational complexity associated with pixel-wise predictions using resolution-preserving blocks \cite{crevnet}. By employing autoencoders and Reversible Predictive Models (RPM), our method minimizes data loss during feature extraction, significantly reduces memory consumption, and enhances processing efficiency. As a result, our lightweight design is well-suited for our specific scenario.

\begin{table}[b]
\vspace{-0.5cm}
\centering
\caption {Simulation Parameters} \label{tab:simu}
\begin{tabular}{|l|p{1.75 cm}||l|l|}
\hline
\textbf{Parameter}           & \textbf{Value}     & \textbf{Parameter}    & \textbf{Value}        \\ \hline
Carrier freq        & 2.4 GHz  &  Deployment scenario & UMa       \\ \hline
Channel model       & 3GPP 38.901   & Subcarrier spacing  & 60 kHz      \\ \hline
System BW    & 100 MHz   & BS max power        & 44 dBm             \\ \hline
Scheduler           & PF, MAX-CQI, DRR & UE tx power     & 23 dBm    \\ \hline
Data rate           &  20,30,60 Mbps    & BS noise figure     & 5 dB \\ \hline
Target BLER         &  1\%  & Numerology index    &  2  \\ \hline
No. of RBs          & 135 & BS Antenna height   & 25 m \\ \hline
UE height           & 1.5m & UE noise figure     & 7 dB \\ \hline

\end{tabular}
\end{table}

\section{Performance Evaluation}\label{sec:perf}

\subsection{Simulation Environment}\label{sec:simu}
We have implemented the 5G NR wireless environment using the Simu5G module, a state-of-the-art platform based on the widely recognized OMNET++ simulator. The scenario consists of multiple XR users receiving service via the Remote Radio Head (RRH), which is connected to the Base Band Unit (BBU) pool through fronthaul connections (refer to Fig. \ref{fig:system_model}). The proposed scheme has been implemented on the Edge server, which is interconnected with the BBU unit. The simulation parameters adhere to the specifications defined by 3GPP and are outlined in Table \ref{tab:simu}. Our reference environment spans a spatial dimension of $250m \times 250m$, with XR users randomly distributed within it. We analyze the results from two perspectives: 1) \textit{Similar users approximation}, where we present network-level performance averaged across similar XR users (those generating at the same frame and data rate), and 2) \textit{Individual user consideration}, where we examine the performance of individual users under different parameters. In all the results, XR users maintain a delay bound of 2.5 ms and a delay reliability threshold of 99\%. The reported results are obtained through the average of 100 independent runs.

\subsection{Traffic Model}\label{traffic_model}
3GPP standardizes the XR traffic model in REL-17 \cite{xr_traffic}. It characterizes XR traffic as pseudo-periodic, where XR devices generate frames following a truncated Gaussian distribution for their sizes. Furthermore, the encoding, compression, and other operations introduce jitter to the frames, which is modeled using another truncated Gaussian distribution. We adopt this standardized XR traffic model to evaluate performance and generate individual frames from the MNIST dataset.

\subsection{Results and Discussions}
\subsubsection{Similar Users Approximation} 
\begin{figure*}[]
\centering
\includegraphics[scale = 0.35]{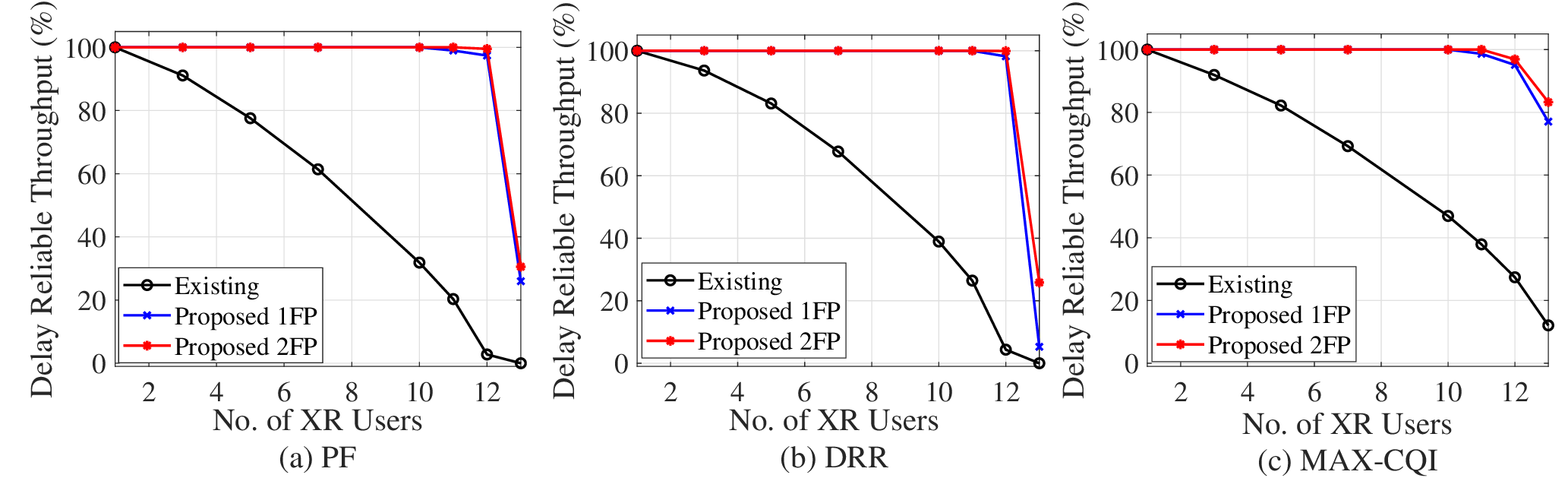}
\caption{{The comparison reveals a significant increase in delay reliable throughput with the proposed one-frame prediction (1FP) or two-frame prediction (2FP) compared to existing schemes (PF, DRR, MAX-CQI) for a frame rate of 120 fps and data rate of 60 Mbps.}}
\label{fig:througput}
\vspace{-0.35cm}
\end{figure*}

We present the network performance analysis of our proposed scheme, showcasing the averaged performances of similar XR users. Fig. \ref{fig:througput} illustrates the performance of \textit{delay reliable throughput}, which denotes the successful transmission percentage of frames within the specified delay constraint. The plot unequivocally indicates a decline in the metric as the number of XR users increases, irrespective of the employed schemes (PF, DRR, or MAX-CQI scheduling). In contrast, the proposed one-frame (1FP, i.e., $p_d^n = 1, \forall n $) and two-frame prediction (2FP) schemes exhibit a minimal effect from XR user counts till the complete exhaustion of the resources, causing a buffer overflow. For instance, while the PF accommodates only one user, our 1FP scheme supports ten users, and the 2FP scheme sustains twelve users, a performance gain surpassing tenfold. Similar outcomes are observed across different schemes, affirming the compatibility of our proposed scheme with existing ones and its capability to deliver substantial performance enhancements. It may be noted that prediction beyond two frames does not yield any tangible gain, which may be because the actual delay never reaches more than a two-frames-relaxed delay budget.

\begin{figure}[b]
\vspace{-0.5cm}
\centering
\includegraphics[scale = 0.43]{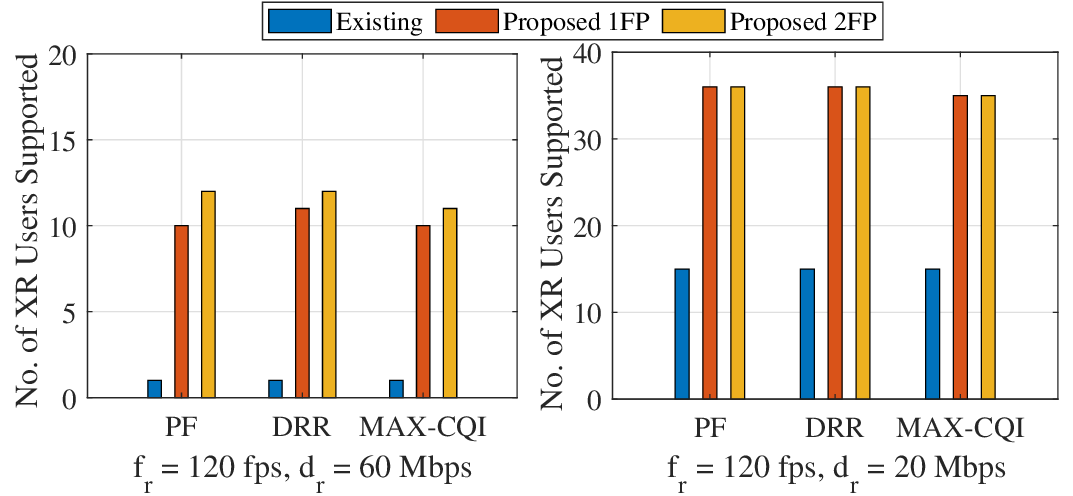}
\caption{{No. of satisfied XR users: the result demonstrates a manifold increase in the number of satisfied XR users.}}
\label{fig:comparison}
\end{figure}

Fig. \ref{fig:comparison} presents a comparative analysis of the supported number of XR users, unequivocally demonstrating the superior performance of our proposed one or two-frame prediction scheme over existing counterparts. Furthermore, the results indicate that existing schemes exhibit relatively better performance when the data rate is decreased while maintaining a fixed frame rate. This is because of the reduction in burstiness resulting from lower data rates, subsequently diminishing the likelihood of exceeding delay-budget thresholds. In this particular case, it is also observed that there is no gain in going beyond one-frame prediction. However, it is imperative to acknowledge that XR traffic is inherently bursty, and in the upcoming future, this is expected to escalate with an even higher data rate. As our proposed scheme excels in managing burstiness and enables support for a substantially higher number of XR users, the provided solution becomes an appropriate choice for upcoming XR scenarios.

\begin{figure}[]
\centering
\includegraphics[scale= 0.35]{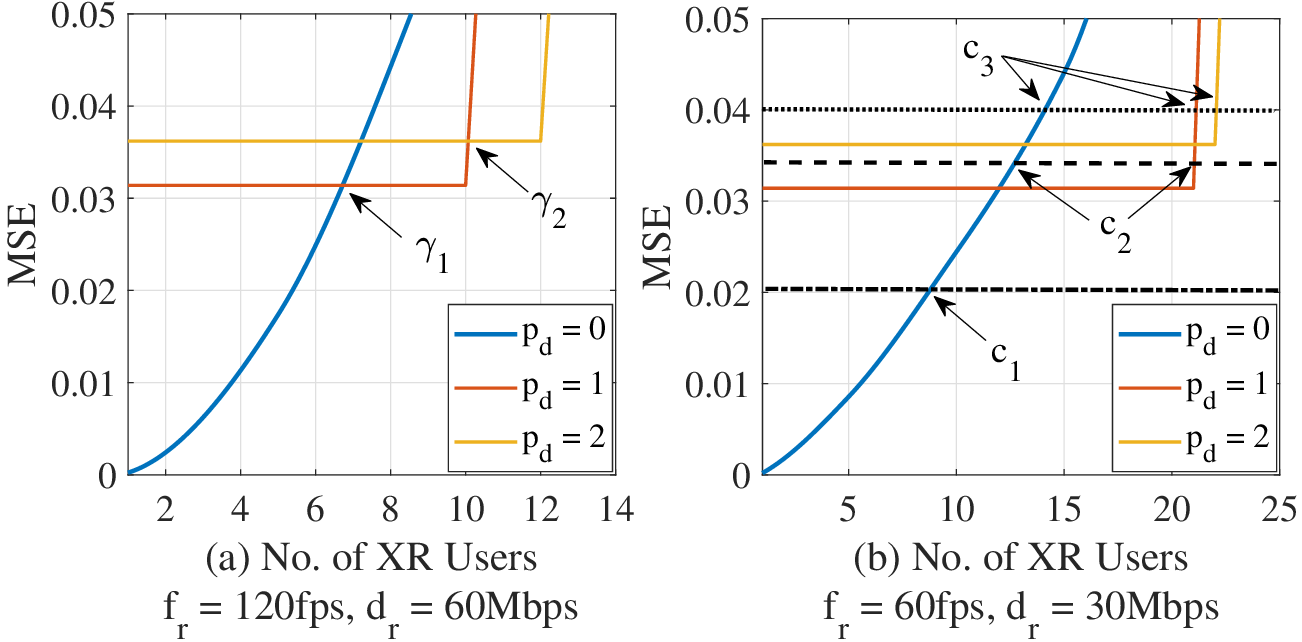}
\caption{{Mean Square Error (MSE) vs. no. of XR users for different data rates and frame rates. Here $p_d \implies p_d^n$ for all $n$.}}
\label{fig:mse_vs_ue}
\vspace{-0.35cm}
\end{figure}

Our proposed method provides significant gains, but a trade-off exists between these gains and prediction errors caused by the prediction scheme employed. We quantify it using the Mean Squared Error (MSE) and show this performance against the number of satisfied XR users in Fig. \ref{fig:mse_vs_ue}. It depicts the MSE of the proposed scheme across different prediction intervals ($p_d$, here $p_d$ implies the same $p_d^n,  \forall n$) for varying frame rates ($f_r$) and data rates ($d_r$), corresponding to different numbers of XR users. Notably, the prediction interval signifies the number of XR frames subjected to prediction. In the plots, $p_d = 0$ indicates prediction solely for XR frames that violate the delay bound. Conversely, for $p_d > 0$, future frames are consistently predicted within the specified interval, and subsequent processing is performed on the predicted data. Fig. \ref{fig:mse_vs_ue} reveals that as the number of XR users increases and more XR frames must adhere to the delay bound, the MSE rises when $p_{d}=0$. Relaxing the delay budget by one frame duration ($p_{d}=1$, i.e., 1FP scheme) allows more users to achieve delay reliability satisfaction up to a specific threshold. At this critical threshold, denoted as $\gamma_{1}$, the $p_{d}=0$ and $p_{d}=1$ curves intersect. This intersection serves as a pivotal network design consideration, indicating that maintaining $p_{d} = 0$ below $\gamma_{1}$ and transitioning to $p_{d}=1$ beyond $\gamma_{1}$ minimizes MSE errors. Enabling a prediction interval of $p_{d}= 2$ (i.e., 2FP scheme) supports an even greater number of users at the expense of increased MSE. However, once the network's resources are fully utilized, the MSE escalates drastically. Similarly, a crossover point $\gamma_{2}$ exists where the $p_{d}=1$ and $p_{d}=2$ curves intersect. To minimize MSE while accommodating a higher number of users, it is advantageous to adopt $p_{d}=1$ below $\gamma_{2}$ and $p_{d}=2$ beyond $\gamma_{2}$.

Figure \ref{fig:mse_vs_ue}b presents a crucial observation that guides the network design process. Three tolerable MSE values are considered and depicted as dotted, dashed, and dash-dotted lines in the plot. The intersection points of these lines with the $p_{d}=0$, $p_{d}=1$, and $p_{d}=2$ curves are marked as $c_1$, $c_2$, and $c_3$. Optimal system performance is attained by selecting the appropriate prediction interval based on the permissible maximum MSE. For instance, if the maximum allowed MSE is set at 0.02, choosing $p_{d}=0$ yields the best outcome, enabling support for approximately eight XR UEs in the reference environment. Likewise, for a maximum allowed MSE of 0.035, maintaining $p_{d}=0$ up to $\gamma_1$ and transitioning to $p_{d}=1$ beyond that point is recommended. By leveraging the crossover point $c_3$, the optimal prediction duration can be determined for a maximum allowed MSE of 0.04. These design insights facilitate achieving the desired system performance level while ensuring the prediction error remains within acceptable bounds.

\subsubsection{Individual User Consideration}
\begin{figure}[]
\centering
\includegraphics[scale = 0.4]{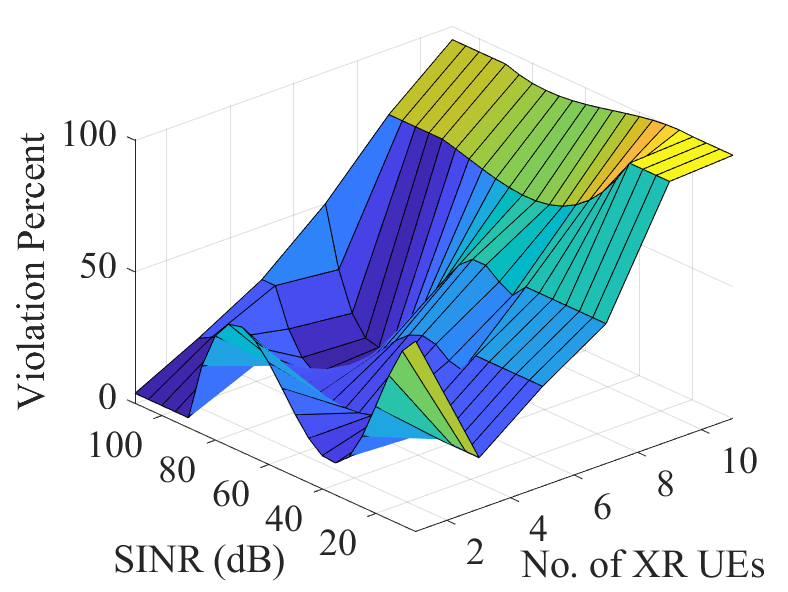}
\caption{{Delay reliability violation percentage of an XR user as a function of its SINR and no. of other XR users.}}
\label{fig:2d}
\vspace{-0.25cm}
\end{figure}

We presented network-level results assuming homogeneous XR users above. Now, we focus on individual users, closely examining their delay reliability violation percentage relative to their SINR values and the number of other XR users. These insights are visualized in Fig. \ref{fig:2d}, where we represent the violation percentage rather than MSE values for enhanced clarity. Our analysis reveals a complex pattern of violations influenced by various factors. In general, when SINR is high, and the number of other XR users is low, the violation percentage remains low. However, as the number of XR users increases, so does the violation percentage. Similarly, decreasing SINR values correspond to higher violation percentages. However, as depicted in the figure, the observed pattern lacks regularity and exhibits complexity. Consequently, clear thresholds for determining prediction intervals cannot be readily established. This highlights the dependence of the violation percentage and, subsequently, prediction interval on multiple factors, including channel conditions, XR user count, and their respective loads. Hence, it is imperative to conduct a comprehensive investigation to identify the factors influencing individual user performance, which can be leveraged as inputs to machine learning models for predicting the duration at the individual user level. Both these aspects will be explored in the future.

\section{Conclusion}\label{sec:conclu}
This letter introduces a novel service provisioning scheme for low-latency XR users in 5G NR, utilizing an AI-based predictor to predict future XR frames. Our scheme effectively extends the delay budget by processing and scheduling the predicted data instead of the actual data, and thereby ensuring the required delay reliability. The results unequivocally demonstrate the effectiveness of our proposed scheme, with a many-fold rise in performance. Moreover, we establish straightforward yet simple network-based rules for zero, one, or two-frame prediction, enabling the satisfaction of more users while maintaining acceptable error levels for homogeneous XR users. However, a closer examination from an individual XR user perspective highlights the complexity of calculating the prediction duration, necessitating further investigation. Future research endeavors may also explore a novel scheduling scheme that capitalizes on the virtual extension of the delay budget, offering even greater performance enhancements for low-latency XR applications.

\bibliography{ref} 
\bibliographystyle{ieeetr}

\end{document}